# Amplification of cascaded down-conversion by reusing photons with a switchable cavity

Alexandre Z. Leger,[*] Samridhi Gambhir, Julien Légère, and Deny R. Hamel[†]

*Département de physique et d'astronomie, Université de Moncton, Moncton, New Brunswick E1A 3E9, Canada*

The ability to efficiently produce and manipulate nonclassical states of light is a critical requirement for the development of quantum optical technologies. In recent years, experiments have demonstrated that cascaded spontaneous parametric down-conversion is a promising approach to implement photon precertification, providing a way to overcome photon transmission losses for quantum communication, as well as to directly produce entangled three-photon states and heralded Bell pairs. However, the low efficiency of this process has so far limited its applicability beyond basic experiments. Here, we propose a scheme to amplify triplet production rates by using a fast switch and a delay loop to reuse photons that fail to convert on the first pass through the cascade's second nonlinear crystal. We construct a theoretical model to predict amplification rates and verify them in an experimental implementation. Our proof-of-concept device increases the rate of detected photon triplets as predicted, demonstrating that the method has the potential to dramatically improve the usefulness of cascaded down-conversion for device-independent quantum communication and entangled state generation.

## I. INTRODUCTION

Device-independent quantum communication aims to implement quantum communication protocols without relying on trust of the quantum devices used to implement them [1,2]. Recently, a significant milestone for the field has been reached as experiments have demonstrated device-independent quantum key distribution [3,4] over short distances. However, extending these experiments to longer distances is not straightforward, as any additional photon transmission losses directly impact the ratio of detected photons, which must remain high to overcome the detection loophole.

Photon precertification is a promising solution to overcome this limitation [5]. By confirming the presence of a photonic qubit before it is detected, precertification makes the transmission losses no longer directly relevant to the detection loophole threshold, with the pertinent efficiency becoming the one between the photon precertification and detection. Photon precertification can be implemented using cascaded spontaneous parametric down-conversion (CSPDC), a process where photons generated from an initial spontaneous parametric down-conversion (SPDC) source are used as a pump for the second. The viability of this approach was demonstrated in a proof-of-principle experiment [6]. However, a key limitation for practical implementation remains the efficiency of the CSPDC process. CSPDC experiments, including those producing three-photon entangled states in the polarization [7,8] and energy-time [9] degrees of freedom, or heralded Bell pairs [7], typically result in very low rates; detected triplet rates in these and other CSPDC experiments have so far been limited to between a few triplets per hour [8–11] and a few hundred per hour [7].

In order to improve these rates and make CSPDC attractive for photon precertification and other practical applications, the most critical bottleneck is the second stage of down-conversion, as any increase to this efficiency will directly enhance the measured rates. In contrast, improving the efficiency of the first stage requires faster detectors to detect the higher rate of photons from the first down-conversion. Increasing the pump power has the same limitation. Moreover, in the context of photon precertification, the generation of photons in the first stage should be considered independent, with the success probability of precertification directly linked to the efficiency of the second state.

It is therefore imperative to explore ways to increase the efficiency of the second stage. One straightforward approach would be to use materials with stronger nonlinearities [12,13]. Another promising technology is thin-film nonlinear devices which use stronger confinement to enhance conversion efficiency. However, these devices have so far reached down-conversion efficiencies on the order of $10^{-8}$ [14,15], whereas commercial periodically poled lithium niobate (PPLN) waveguides such as the one used in this work already achieve efficiencies greater than $10^{-6}$ [10]. Another option could be to place the second SPDC stage inside a cavity, which is a commonly used approach to enhance SPDC efficiency [16]. However, the benefit of such an approach would be limited if the acceptance bandwidth of this cavity is smaller than that of the photons produced in the first stage. While such a cavity could be an attractive option if the first stage consists of an atomic source [17], or of an SPDC source which is also placed in its own cavity with both cavities locked together, it is not immediately implementable if the first stage is a regular SPDC process.

*eal2753@umoncton.ca
†deny.hamel@umoncton.ca

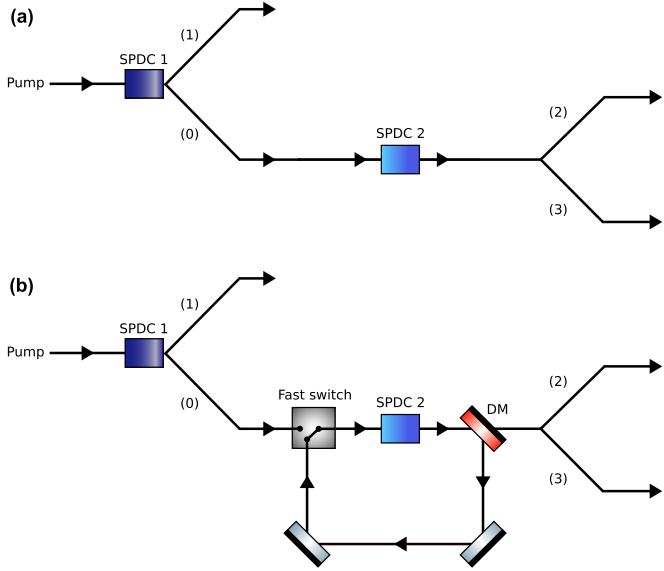

FIG. 1. Comparison of traditional CSPDC and our amplified CSPDC scheme. (a) A regular CSPDC scheme for producing time-correlated photon triplets. (b) An enhanced CSPDC scheme for producing time-correlated photon triplets with amplification by an active loop. The first pair is created at SPDC 1. The fast switch is shown to be in the closed position. With the right timing, triggered by a heralded photon pair or a pulsed laser, the switch is opened just before photon (0) arrives. Photon (0) is given a chance at down-conversion at SPDC 2, then a dichroic mirror (DM) reflects it back into the loop. Until another photon pair arrives from SPDC 1, this photon will be given many chances to interact with SPDC 2 to create a second pair and thus a triplet.

In this work, we propose a different approach which aims to get the benefits of cavity enhancement without the typical bandwidth constraints of time-independent cavities. One way around these limitations is through the use of time-dependent components [18,19]. Here, we make use of the fact that arrival times of photons from the first down-conversion are known in order to capture them inside a switchable cavity consisting of a delay loop and fast switch. This approach is similar to setups producing heralded single photons using temporal multiplexing based on storage cavities or loops [20], but with the addition of a second nonlinear crystal within the storage cavity. Photons that fail to convert on the first pass through this second nonlinear crystal are therefore provided with multiple additional chances of down-conversion, thus amplifying the production of triplets. To validate the concept, we develop a theoretical model to investigate the expected amplification from this method and provide experimental results from a proof-of-concept device.

## II. THEORY

Our scheme, shown in Fig. 1(b), consists of a CSPDC device with an active loop controlled by a fast switch. This optical switch should be open when a photon in mode (0) is expected to arrive. The timing can be determined either using the detection of its partner photon in mode (1) or, in the case of a pulsed laser, with a periodic signal synchronized with the pump. It is important that the switch is fast enough to be in a closed state before the photon completes the loop. Otherwise, the photon will escape during its second pass instead of being trapped. To determine the quantum state produced by this setup, we follow the approach of Ref. [10], modeling CSPDC as a combination of two Hamiltonians: $\hat{H}_1 = g_1\alpha(\hat{a}_0^\dagger \hat{a}_1^\dagger + \text{H.c.})$ and $\hat{H}_2 = g_2(\hat{a}_0 \hat{a}_2^\dagger \hat{a}_3^\dagger + \text{H.c.})$, where the $g_i$ represent the respective coupling strength of each interaction, $\alpha$ is the classical amplitude of the pump laser, $a_i^\dagger$ are creation operators in the respective modes, and H.c. represents the Hermitian conjugate. Here, instead of writing the transformation as $\hat{U}_2\hat{U}_1 = \exp(-i\hat{H}_2)\exp(-i\hat{H}_1)$ as in regular CSPDC, $\hat{U}_2$ should be applied several times, adding in a time-delay and loss caused by the loop between each interaction. We therefore model the transformation after $j$ passes as $\hat{U} = (\hat{U}_3\hat{U}_2)^j\hat{U}_1$, where $\hat{U}_3$ encapsulates the delay $\tau$ and limited transmission $\beta$ of the loop. Modeling the loss as a beam-splitter transformation, we write the transformation effected by $\hat{U}_3$ on mode 0 as

$$\hat{a}_{0,t}^\dagger \to \sqrt{\beta}\hat{a}_{0,t+\tau}^\dagger + \sqrt{1-\beta}\hat{a}_{\text{loss},t+\tau}^\dagger, \quad (1)$$

where the new subscript indicates the timing $t$ of the creation operators. Applying $\hat{U}$ to the vacuum state, and keeping only terms at most quadratic in $g_i$ which produce at least one photon in modes 1, 2, and 3, the resulting state is

$$|\Psi\rangle \approx -g_1 g_2 \alpha \sum_{k=0}^{j}(\sqrt{\beta})^k |0\rangle_0 |1\rangle_1 |1\rangle_{2,k\tau} |1\rangle_{3,k\tau}. \quad (2)$$

By neglecting higher-order terms of $g_i$, this expression ignores multipair events produced from each down-conversion. The zeroth-order term in the sum is identical to the case of regular CSPDC, with subsequent terms only differing by a reduced amplitude and an added delay on photons in modes 2 and 3.

We now consider the enhancement of rates expected from this approach. Without the optical switch, or by keeping it in its "open" state, the expected rate of produced triplets is given by

$$T_0 = R_1 P_{\text{SPDC}}, \quad (3)$$

with $R_1$ being the rate of photons in mode (1), and $P_{\text{SPDC}}$ being the probability of down-conversion at SPDC 2, respectively proportional to $|\alpha g_1|^2$ and $|g_2|^2$ in the low-pump regime described in Eq. (2). With the addition of the switch, the photon in mode (0) is trapped in the loop and gets multiple passes through SPDC 2. The contribution to the overall triplet rate from the $k$th pass is given by

$$T_k = \eta R_1 P_{\text{SPDC}}(1 - P_{\text{SPDC}})^k \beta^k, \quad (4)$$

where $\eta$ is the efficiency of the switch and $\beta$ is the loop transmission efficiency, defined as the transmission probability after one full loop. We define the amplification factor $A$ as the ratio of the sum of contributions from each $T_k$ term with the unamplified rate:

$$A = \eta \sum_{k=0}^{\infty}[\beta(1-P_{\text{SPDC}})]^k. \quad (5)$$

Equation (5) is a geometric series and the expected amplification from this scheme is thus

$$A = \frac{1}{1 - \beta(1 - P_{\text{SPDC}})}. \quad (6)$$

This first model is valid as long as the switch is never activated while a photon is still in the loop. For a more accurate calculation, we must take into account the effect of the switch being triggered while another photon is present. If the switch is fast enough, it could be possible to have more than one photon inside the loop at a time. However, this would lead to complications in discriminating triplets from one another. Instead, we assume that the switch is always opened long enough such that its activation for a new photon will eject any photon already inside the loop. Here, we consider the specific case of a continuous pump with triggers to the switch coming from detections of photon (1). Since SPDC will lead to probabilistic detection events at detector 1, we calculate the probability that no photons are detected during one trip around the loop. Detection events from SPDC could follow either a Poissonian or thermal distribution [21]. However, given the limited detection efficiency of photon 1, we expect the distribution to become Poissonian in either case [22]. We therefore use the Poisson distribution, $\text{Pois}(x, \lambda) = \frac{\lambda^x e^{-\lambda}}{x!}$, evaluated at $x = 0$, to calculate the probability of no photons being detected in mode (1) within the loop time $t_L$:

$$\text{Pois}(x = 0, \lambda = R_1 t_L) = \exp(-R_1 t_L), \quad (7)$$

where $\lambda$ is the expected number of photons in mode (0) in the time interval $t_L$. Note that if the detection efficiency would be sufficiently high for the detections to follow a thermal distribution, Eq. (7) would still be a good approximation as long as $R_1 t_L \ll 1$. We thus find that the active switch gives an amplification of

$$A = \eta \sum_{k=0}^{\infty} \left[ \frac{\beta(1 - P_{\text{SPDC}})}{\exp(R_1 t_L)} \right]^k. \quad (8)$$

Typically, the single-photon detectors have a certain dead time $\tau$ when no detections can be made, therefore no new triggers are registered during this time. We define the quantity $j = \lfloor t_L / \tau \rfloor$, which is a floor function that gives the integer number of passes in which the photon that is trapped in the loop cannot be affected by any new photon pairs. The contribution from the first $j$ passes will follow Eq. (5), whereas for passes beyond the value of $j$, the Poisson term must be taken into account as in Eq. (8). The more complete model for the amplification is thus given by

$$A = \eta \sum_{k=0}^{j} [\beta(1 - P_{\text{SPDC}})]^k + \eta \sum_{k=j+1}^{\infty} \left[ \frac{\beta(1 - P_{\text{SPDC}})}{\exp(R_1 t_L)} \right]^k. \quad (9)$$

As can be seen in Fig. 2, at low values of alpha the amplification is modest and there is almost no dependence on the SPDC 1 rate. This is because most photons in the loop will be lost before getting enough passes to be converted or will be ejected by a subsequent photon from SPDC 1. However, at higher values of loop efficiency the amplification can be significant: for a heralding rate of around 1 million counts per second, which is reasonable for a commercially available single-photon avalanche diode, a loop efficiency of 0.93 is sufficient to provide an order of magnitude amplification. For higher count rates, ejection of photons from the loop becomes significant, so a shorter loop time would become beneficial.

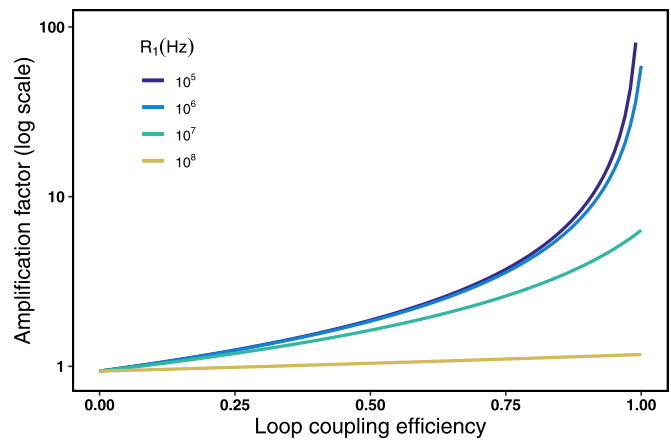

FIG. 2. Dependence of the amplification factor $A$ [Eq. (9)] on the loop coupling efficiency $\beta$, for different mode (1) photon rates. The parameters used in these predictions, which are similar to those in our experiment, are $\eta = 0.94$, $P_{\text{SPDC}} = 1.5 \times 10^{-6}$, $\tau = 45$ ns, and $t_L = 23$ ns. As the mode (1) photon rate increases, the theoretical maximum amplification is reduced. This is due to new photons arriving and ejecting the already present photon from the loop, reducing the number of passes that each photon has in the second SPDC crystal.

## III. EXPERIMENT

To demonstrate the validity of the enhancement model, we implement the scheme experimentally as shown in Fig. 3. The cascade starts with a 405-nm continuous-wave (CW) laser (Toptica Topmode) pump in a periodically poled potassium triphosphate (PPKTP) crystal (Raicol) which produces photon pairs at 775 nm and 849 nm. The 849-nm photons are detected first, while the 775-nm photons pass through a fiber optic delay line before being sent to the fast switch. To implement the fast switch, a Pockels cell (PC) is combined with a polarizing beam splitter (PBS) [23,24]. The PC is tuned as a half-wave plate so that the loop is open at high voltage and closed when the voltage is off. After the PBS and before passing through the PC, the 775-nm photons are horizontally polarized. The trigger signal from the single photon avalanche detectors (SPADs) (Excelitas SPCM-AQ4C) detecting 849-nm photon switches on the PC and the polarization of the 775-nm beam is changed from horizontally polarized to vertically polarized. This is the required pump polarization for type-0 SPDC at the second nonlinear crystal, a PPLN waveguide (HC Photonics).

The timing of the PC is set to allow sufficient time for it to reach half-wave voltage before the 775-nm photon arrives. The 775-nm photons which do not down-convert within the nonlinear crystal are reflected by a DM towards the PBS. Since these photons are now vertically polarized, they are reflected by the PBS and sent to the PC. The time in which the PC is on after the photon enters the loop is sufficiently low such that by the time the vertically polarized 775-nm photons reach the PC, it is turned off. This ensures that the polarization is not changed a second time and the photons can stay in the loop until they are down-converted or absorbed.

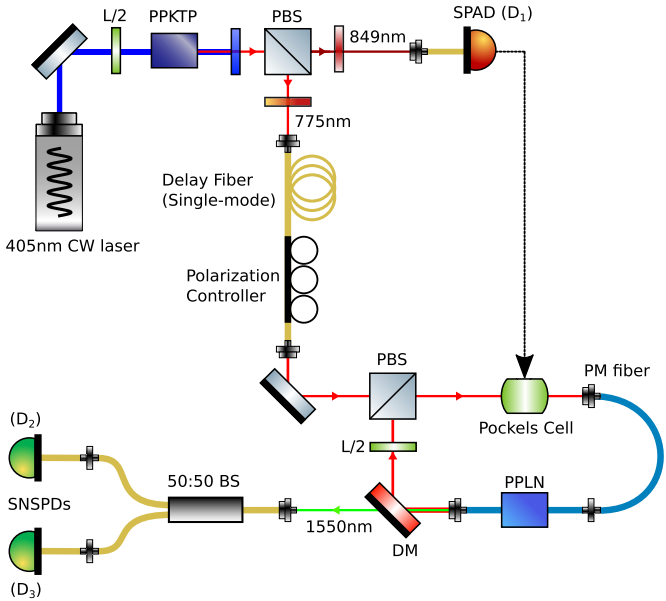

FIG. 3. A 405-nm pump beam, with polarization set by a half-wave plate (L/2), is incident on a PPKTP crystal. Type-2 SPDC occurs at the PPKTP crystal, which produces two orthogonally polarized photon beams of wavelengths 849 nm and 775 nm. These two overlapping beams are separated using a polarizing beam splitter (PBS), followed by appropriate band-pass filters for removing the remaining pump beam. The 775-nm photon is sent to a second stage of SPDC for the creation of a second time-correlated pair. Once the polarization has been modified by the Pockels cell, the photons are coupled into a polarization maintaining (PM) fiber coupled to a PPLN waveguide. The output of the waveguide is coupled back into free space and is incident on a dichroic mirror (DM) which reflects the 775-nm light back into the loop and transmits the 1550-nm light towards the 50:50 beam splitter (BS) and then to the detectors. The signals from the three detectors are sent to a logic correlation unit for triplet detection. The half-wave plate (HWP) in the loop is used to control the loop coupling efficiency in order to test the model.

Photon pairs produced inside the PPLN waveguide are separated probabilistically before being detected by superconducting nanowire single-photon detectors (SNSPDs) (Photon Spot). The time stamps of each triplet detection are recorded using a time tagging unit with 156.25-ps time resolution (UQDevices Logic 16), with subsequent events being labeled as a triplet if detections are recorded at all three detectors with a maximum delay between each detection of 95.9375 ns.

## IV. RESULTS

With a 4.3-mW pump incident on the PPKTP crystal, 330 000 coincidences per second, within a coincidence window of 1 ns, are detected between the 775-nm and 849-nm beams at the SPADs. Compared to a half-wave plate, the PC converts the polarization of $\eta_{PC} = 94\% \pm 3\%$ of the 775-nm photons. This loss of efficiency is mainly due to photons arriving within 100 ns of a previous detection, which are ignored by the PC driver to protect the high-voltage circuit. The PPLN waveguide used for cascaded SPDC has a down-conversion efficiency of $P_{SPDC} \approx 10^{-6}$. The combined detection and coupling efficiency for the 1550-nm photons after the beam splitter is 0.156, measured as the ratio of coincidence and single photon detections.

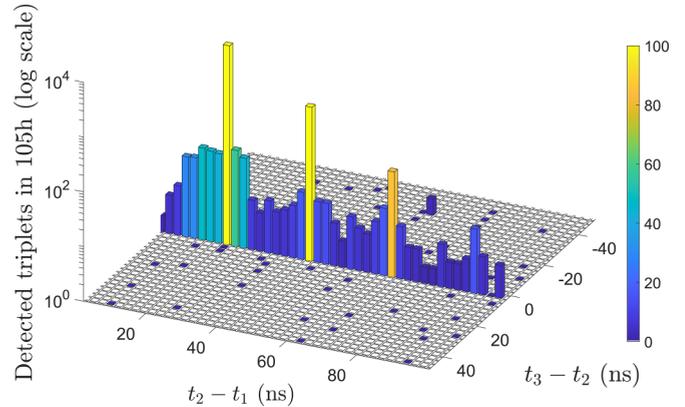

FIG. 4. A three-dimensional histogram of the triplet counts for our maximum achievable loop coupling efficiency $\beta_0$. The $t_2 - t_1$ axis is the time difference between a 849-nm trigger photon at $D_1$ and a 1550-nm photon at $D_2$. The $t_3 - t_2$ axis is the time difference between 1550-nm detections at $D_2$ and $D_3$. The color bar is set to saturate at 100 detections to maximize the visibility of the four peaks.

Figure 4 shows the time distribution of all photon triplets measured during 105 h with the loop coupling efficiency at its maximum value of 18%. Most of the triplets lie on the $t_3 - t_2 = 0$ line, since the 1550-nm photons are created simultaneously and travel equal distances. From left to right, the first and largest peak in the histogram represents photons that down-converted on the first pass through the PPLN waveguide (SPDC 2). The second peak corresponds to triplets generated after one additional pass through the PPLN waveguide, and so on. The time difference between the two consecutive peaks is the time it takes the 775 photons to traverse the loop once, 23 ns. Most accidentals in our experiment, visible as triplets on the fixed $t_3 - t_2 = 0$ line but not within these main peaks, are therefore caused by a real 1550-nm pair being detected along with an uncorrelated 849-nm photon. These events, corresponding to double pairs from the first SPDC which are ignored in Eq. (2), scale quadratically with the pump power and would impact the fidelity of entangled states produced with this method. Based on this measurement, we find an average triplet rate of $48.9 \pm 0.7$ triplets/h. As a first estimate of the amplification, we can compare this value with a 35-h measurement performed with the half-wave plate in the delay loop rotated by 45°, which minimizes the loop transmission to $\beta = 8.1 \times 10^{-3}$. From this measurement, we obtain a rate of $39.5 \pm 1.1$ triplets/h, which would imply an increase to the raw triplet rate by a factor of $1.24 \pm 0.04$ with the active switching. Such a comparison is not ideal, however, as setup fluctuations in each long measurement mean that the raw triplet rates could vary too much to be a reliable measure of amplification. To address this, we instead infer the experimental amplification $A_{exp}$ by comparing the relative size

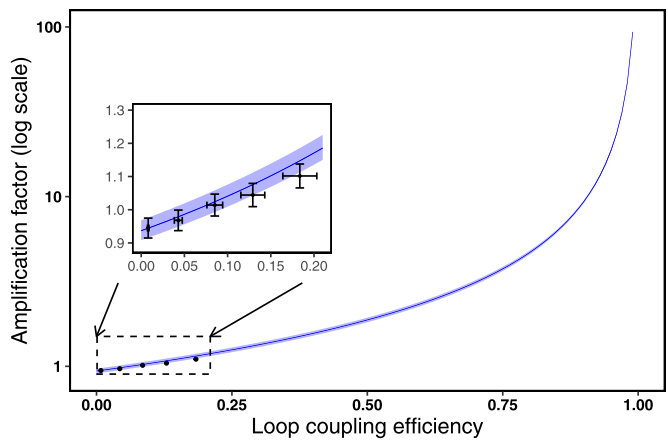

FIG. 5. Comparison of measured amplification with model prediction. Each experimental data point corresponds to a 35-h measurement, except for the $\beta = 0.18$ data point, which was measured for 105 h. The error bars are only shown on the inset. For amplification, they are calculated assuming Poissonian statistics on the number of counts in each peak combined with the uncertainty in $\eta_{PC}$. The uncertainty in $\beta$ are estimated based on observed variations in fiber coupling efficiencies in the setup. The uncertainty band for the model prediction come from uncertainties in the values of $\eta_{PC}$ and $R_1$, given by $\eta_{PC} = 0.94 \pm 0.03$ and $R_1 = (1.52 \pm 0.03) \times 10^6$ counts/s.

of the peaks in the timing histograms. The amplification is calculated as

$$A_{\exp} = \eta_{PC}\left(1 + \frac{P_2 + P_3 + P_4}{P_1}\right), \qquad (10)$$

where $P_i$ is the triplet count on the $i$th peak. We include $\eta_{PC} = 0.94 \pm 0.03$ in the calculation in order to quantify whether any advantage coming from the additional peaks is enough to compensate for the PC not being perfectly efficient. We measure $A_{\exp}$ for various values of the loop coupling efficiency $\beta$ in order to verify the model.

The results of these measurements are shown in Fig. 5 along with the predictions of the model. For low values of $\beta$, the amplification is initially less than 1, since the gain from multiple loops is not enough to compensate for the imperfect switching of the Pockels cell. Then, once $\beta > 0.07$, we see a gain and the switchable cavity becomes beneficial. At the maximum value of $\beta = 0.18 \pm 0.02$, we find an net amplification of $A_{\exp} = 1.10 \pm 0.04$. Ignoring the Pockels cell efficiency, this corresponds to an amplification of $1.18 \pm 0.04$. This is consistent with the preliminary estimate of $1.24 \pm 0.04$, although we expect the measurement based on the relative sizes of the peaks to be more reliable, as explained earlier.

## V. DISCUSSION

While the experimental results are in agreement with our model, further improvements are required for the switchable cavity to provide an appreciable improvement to cascaded down-conversion for practical application. First, a drawback of our implementation is that polarization is employed to implement the switch, which prevents the use of this same degree of freedom to encode quantum information on the photons. This scheme is therefore not immediately compatible with three-photon polarization entanglement or with the precertification of polarization qubits. It is, however, appropriate with qubits encoded in other degrees of freedom—for instance, with ultrafast time bins [25]. Alternatively, the scheme could be implemented using different switching methods which do not exploit polarization [26], thus enabling compatibility with this degree of freedom.

Furthermore, in order to reach a regime where the rate enhancement justifies the added complexity, it will be necessary to increase the loop coupling efficiency. Coupling efficiencies of over 85% into single-mode fibers have been reached out of nonlinear crystal waveguides [4,27], so a similar coupling efficiency should be achievable from a waveguide to itself. With a comparable loop coupling efficiency, our model predicts that an amplification factor of 6 should be accessible with current technology.

Even higher loop efficiency would likely be possible with bulk nonlinear crystals to have a fully free-space cavity, in which case the limiting factors would likely become losses from the optical components. However, it is unclear if any gains from reusing photons would be enough in this case to compensate for the loss in down-conversion efficiency caused by moving away from waveguide devices.

Alternatively, the scheme may benefit from being implemented entirely as an integrated photonic circuit. The production of photon triplets from cascaded down-conversion has already been demonstrated to be possible on an integrated chip [11]. The Pockels effect has also been shown to be viable on integrated devices [28,29]. It may therefore be possible to integrate this setup, including the second down-conversion stage and the switchable cavity, on a photonic chip, which could reduce coupling losses between the fast switch and the down-conversion stage. Another advantage of employing integrated circuits is the possibility of smaller loop lengths. This is beneficial at high loop coupling efficiency, since in this regime it becomes necessary to allow for a large number of passes to happen before the switch is reactivated.

With such increases to the loop coupling efficiency, reusing photons with a switchable cavity will become a viable approach to increase the production rates of photon triplets from CSPDC by more than an order of magnitude. This will make it attractive to optical quantum computing schemes which rely on either three-photon entangled states [30] or heralded Bell pairs [31] as a resource. If our amplification scheme is accompanied with a high detection efficiency of photons from the second down-conversion, the performance of photon precertification will be improved, paving the way towards long-distance device-independent quantum communication.

## ACKNOWLEDGMENTS

We acknowledge the support of the Natural Sciences and Engineering Research Council of Canada, the Canada Foundation for Innovation, Canada Research Chairs, and the New Brunswick Innovation Foundation. The authors thank Z. M. E. Chaisson for contributions to the data collection software.

[1] J. Barrett, L. Hardy, and A. Kent, No Signaling and Quantum Key Distribution, Phys. Rev. Lett. **95**, 010503 (2005).

[2] A. Acín, N. Brunner, N. Gisin, S. Massar, S. Pironio, and V. Scarani, Device-Independent Security of Quantum Cryptography against Collective Attacks, Phys. Rev. Lett. **98**, 230501 (2007).

[3] D. P. Nadlinger, P. Drmota, B. C. Nichol, G. Araneda, D. Main, R. Srinivas, D. M. Lucas, C. J. Ballance, K. Ivanov, E. Y.-Z. Tan, P. Sekatski, R. L. Urbanke, R. Renner, N. Sangouard, and J.-D. Bancal, Experimental quantum key distribution certified by Bell's theorem, Nature (London) **607**, 682 (2022).

[4] W. Zhang, T. van Leent, K. Redeker, R. Garthoff, R. Schwonnek, F. Fertig, S. Eppelt, W. Rosenfeld, V. Scarani, C. C.-W. Lim, and H. Weinfurter, A device-independent quantum key distribution system for distant users, Nature (London) **607**, 687 (2022).

[5] A. Cabello and F. Sciarrino, Loophole-Free Bell Test Based on Local Precertification of Photon's Presence, Phys. Rev. X **2**, 021010 (2012).

[6] E. Meyer-Scott, D. McCloskey, K. Gołos, J. Z. Salvail, K. A. G. Fisher, D. R. Hamel, A. Cabello, K. J. Resch, and T. Jennewein, Certifying the Presence of a Photonic Qubit by Splitting It in Two, Phys. Rev. Lett. **116**, 070501 (2016).

[7] D. R. Hamel, L. K. Shalm, H. Hübel, A. J. Miller, F. Marsili, V. B. Verma, R. P. Mirin, S. W. Nam, K. J. Resch, and T. Jennewein, Direct generation of three-photon polarization entanglement, Nat. Photonics **8**, 801 (2014).

[8] Z. M. E. Chaisson, P. F. Poitras, M. Richard, Y. Castonguay-Page, P.-H. Glinel, V. Landry, and D. R. Hamel, Phase-stable source of high-quality three-photon polarization entanglement by cascaded down-conversion, Phys. Rev. A **105**, 063705 (2022).

[9] S. Agne, T. Kauten, J. Jin, E. Meyer-Scott, J. Z. Salvail, D. R. Hamel, K. J. Resch, G. Weihs, and T. Jennewein, Observation of Genuine Three-Photon Interference, Phys. Rev. Lett. **118**, 153602 (2017).

[10] H. Hübel, D. R. Hamel, A. Fedrizzi, S. Ramelow, K. J. Resch, and T. Jennewein, Direct generation of photon triplets using cascaded photon-pair sources, Nature (London) **466**, 601 (2010).

[11] S. Krapick, B. Brecht, H. Herrmann, V. Quiring, and C. Silberhorn, On-chip generation of photon-triplet states, Opt. Express **24**, 2836 (2016).

[12] X. Guo, C.-l. Zou, C. Schuck, H. Jung, R. Cheng, and H. X. Tang, Parametric down-conversion photon-pair source on a nanophotonic chip, Light: Sci. Appl. **6**, e16249 (2017).

[13] E. J. Stanton, J. Chiles, N. Nader, G. Moody, G. Moody, N. Volet, L. Chang, J. E. Bowers, S. W. Nam, and R. P. Mirin, Efficient second harmonic generation in nanophotonic GaAs-on-insulator waveguides, Opt. Express **28**, 9521 (2020).

[14] J. Zhao, C. Ma, M. Rüsing, and S. Mookherjea, High Quality Entangled Photon Pair Generation in Periodically Poled Thin-Film Lithium Niobate Waveguides, Phys. Rev. Lett. **124**, 163603 (2020).

[15] C. J. Xin, J. Mishra, C. Chen, D. Zhu, A. Shams-Ansari, C. Langrock, N. Sinclair, F. N. C. Wong, M. M. Fejer, and M. Lončar, Spectrally separable photon-pair generation in dispersion engineered thin-film lithium niobate, Opt. Lett. **47**, 2830 (2022).

[16] O. Slattery, L. Ma, K. Zong, and X. Tang, Background and review of cavity-enhanced spontaneous parametric down-conversion, J. Res. Natl. Inst. Stand. **124**, 124019 (2019).

[17] D.-S. Ding, W. Zhang, S. Shi, Z.-Y. Zhou, Y. Li, B.-S. Shi, and G.-C. Guo, Hybrid-cascaded generation of tripartite telecom photons using an atomic ensemble and a nonlinear waveguide, Optica **2**, 642 (2015).

[18] S. A. Mann, D. L. Sounas, and A. Alù, Nonreciprocal cavities and the time–bandwidth limit, Optica **6**, 104 (2019).

[19] I. Cardea, D. Grassani, S. J. Fabbri, J. Upham, R. W. Boyd, H. Altug, S. A. Schulz, K. L. Tsakmakidis, and C.-S. Brès, Arbitrarily high time bandwidth performance in a nonreciprocal optical resonator with broken time invariance, Sci. Rep. **10**, 15752 (2020).

[20] E. Meyer-Scott, C. Silberhorn, and A. Migdall, Single-photon sources: Approaching the ideal through multiplexing, Rev. Sci. Instrum. **91**, 041101 (2020).

[21] H. Takesue and K. Shimizu, Effects of multiple pairs on visibility measurements of entangled photons generated by spontaneous parametric processes, Opt. Commun. **283**, 276 (2010).

[22] M. Fox, *Quantum Optics: An Introduction*, Oxford Master Series in Physics (Oxford University Press, Oxford, 2006).

[23] T. B. Pittman, B. C. Jacobs, and J. D. Franson, Single photons on pseudodemand from stored parametric down-conversion, Phys. Rev. A **66**, 042303 (2002).

[24] F. Kaneda, B. G. Christensen, J. J. Wong, H. S. Park, K. T. McCusker, and P. G. Kwiat, Time-multiplexed heralded single-photon source, Optica **2**, 1010 (2015).

[25] F. Bouchard, D. England, P. J. Bustard, K. Heshami, and B. Sussman, Quantum communication with ultrafast time-bin qubits, PRX Quantum **3**, 010332 (2022).

[26] X.-s. Ma, S. Zotter, J. Kofler, T. Jennewein, and A. Zeilinger, Experimental generation of single photons via active multiplexing, Phys. Rev. A **83**, 043814 (2011).

[27] A. B. U'Ren, C. Silberhorn, K. Banaszek, and I. A. Walmsley, Efficient Conditional Preparation of High-Fidelity Single Photon States for Fiber-Optic Quantum Networks, Phys. Rev. Lett. **93**, 093601 (2004).

[28] B. Chmielak, M. Waldow, C. Matheisen, C. Ripperda, J. Bolten, T. Wahlbrink, M. Nagel, F. Merget, and H. Kurz, Pockels effect based fully integrated, strained silicon electro-optic modulator, Opt. Express **19**, 17212 (2011).

[29] S. Abel, F. Eltes, J. E. Ortmann, A. Messner, P. Castera, T. Wagner, D. Urbonas, A. Rosa, A. M. Gutierrez, D. Tulli, P. Ma, B. Baeuerle, A. Josten, W. Heni, D. Caimi, L. Czornomaz, A. A. Demkov, J. Leuthold, P. Sanchis, and J. Fompeyrine, Large Pockels effect in micro- and nanostructured barium titanate integrated on silicon, Nat. Mater. **18**, 42 (2019).

[30] M. Gimeno-Segovia, P. Shadbolt, D. E. Browne, and T. Rudolph, From Three-Photon Greenberger-Horne-Zeilinger States to Ballistic Universal Quantum Computation, Phys. Rev. Lett. **115**, 020502 (2015).

[31] D. E. Browne and T. Rudolph, Resource-Efficient Linear Optical Quantum Computation, Phys. Rev. Lett. **95**, 010501 (2005).